\documentstyle[epsfig]{aipproc}

\def\eset{e_{\rm s}/e_{\rm k}}
\def\etal{{\it et al.}}

\def\Fs{\bf F_{\rm s}}

\def\game{\gamma_{\rm e}}

\def\o{{\rm o}}
\def\ppm{{\it ppm}}

\def\u{{\bf u}}
\def\wp{{\bf \omega}_{\rm p}}

\begin{document}
\title{Highly Compressible MHD Turbulence and Gravitational Collapse}
\thanks{This work has received
partial finantial support from grants UNAM/CRAY grant SC-002395, UNAM-DGAPA
IN105295, and a joint CONACYT-CNRS grant.}

\author{E. V\'azquez-Semadeni$^*$, T. Passot$^{\dagger}$  and A. Pouquet
$^{\dagger}$}
\address{
$^*$Instituto de Astronom\'\i a, UNAM
Apdo.\ Postal 70-264, M\'exico, D.F. 04510
$^{\dagger}$Observatoire de la C\^ote d'Azur, B.P.\ 4229, 06304, Nice Cedex 4,
France\\
}

\lefthead{VAZQUEZ-SEMADENI ET AL.}
\righthead{HIGHLY COMPRESSIBLE MHD TURBULENCE}
\maketitle

\begin{abstract}
We investigate the properties of highly
compressible turbulence and its ability to produce self-gravitating
structures. The compressibility is parameterized
by an effective polytropic exponent $\game$.
In the limit of small $\game$, the density jump at
shocks is shown to be of the order of $e^{M^2}$, and the production
of vorticity by the nonlinear terms appears to be negligible.
%
%
In the presence of self-gravity, we suggest that turbulence can produce bound
structures for $\game < 2(1-1/n)$, where $n$ is the typical
dimensionality of the turbulent compressions.
We show, by means of numerical simulations, 
that, for sufficiently small $\game$,
small-scale turbulent density fluctuations eventually
collapse even though the medium is globally stable. This result is
preserved in the presence of a magnetic field for supercritical
mass-to-flux ratios.
\end{abstract}

\section*{Introduction}
%
%
%
%

In this paper we present a brief discussion on
the dynamical properties of highly
compressible turbulence, with and without self-gravity. A full-length
discussion can be found in V\'azquez-Semadeni \etal\ 
(1996, hereafter Paper I). The
compressibility of the medium is parameterized by an effective
polytropic exponent $\game$ arising from balance between heating and cooling
processes (Elmegreen 1991; V\'azquez-Semadeni \etal\ 1995; Passot \etal\ 1995),
such that the pressure $P$ and the density
$\rho$ are related by $P \propto \rho^{\game}$. We
consider the
cases where $\game$ is either constant, or has a piecewise
density dependence (a ``piecewise polytropic model'' or \ppm). We also
consider fully thermodynamic cases.
Most of the results are based on 
numerical calculations which solve the full MHD equations in two or three
dimensions, at resolutions of $128^2$ or $64^3$, respectively (Paper I).
 
\section*{No Self-Gravity}
 
\subsection*{Density Structures}
 
It can be easily shown that the density jump $X \equiv \rho_2/\rho_1$
across a shock in a barotropic gas of index $\game$ satisfies:
\begin{equation}
X^{1+\game} - (1+ \game M^2)X + \game M^2=0,
\end{equation}
where $M$ is the upstream Mach number.
 

In the isothermal case where $\game = 1$, $X =M^2$, while for $0 <
\game \ll 1$, $X \sim e^{M^2}$. This implies that for small $\game$ :
 
\noindent
i) The density jump is much larger than in the isothermal case.
 
\noindent
ii) Less than supersonic motions (with respect to the isothermal sound speed)
are sufficient for producing large density fluctuations.
 
%
 
\subsection*{Vorticity evolution}
 
From the momentum conservation equation (Paper I), we can derive the evolution
equation for the {\it potential vorticity} $\wp \equiv \omega/\rho
\equiv \nabla \times \u /\rho$ :
\begin{equation}
{\partial \wp \over \partial t} + \u \cdot \nabla \wp = \wp \cdot
\nabla \u + \nabla \times \Fs + \nabla P \times \nabla \rho/\rho^3
\end{equation}
where $\Fs$ is the solenoidal (or rotational)
part of the turbulent energy sources.
 
While compressive motions can easily be generated from vortical
motions,  only the ``vortex stretching'' term $\wp \cdot \nabla \u$
and the ``baroclinic'' term (nonzero only in the fully thermodynamic
case) are available as nonlinear sources of $\wp$. How effective are they?
 
\begin{figure}
\centerline{\epsfig{file=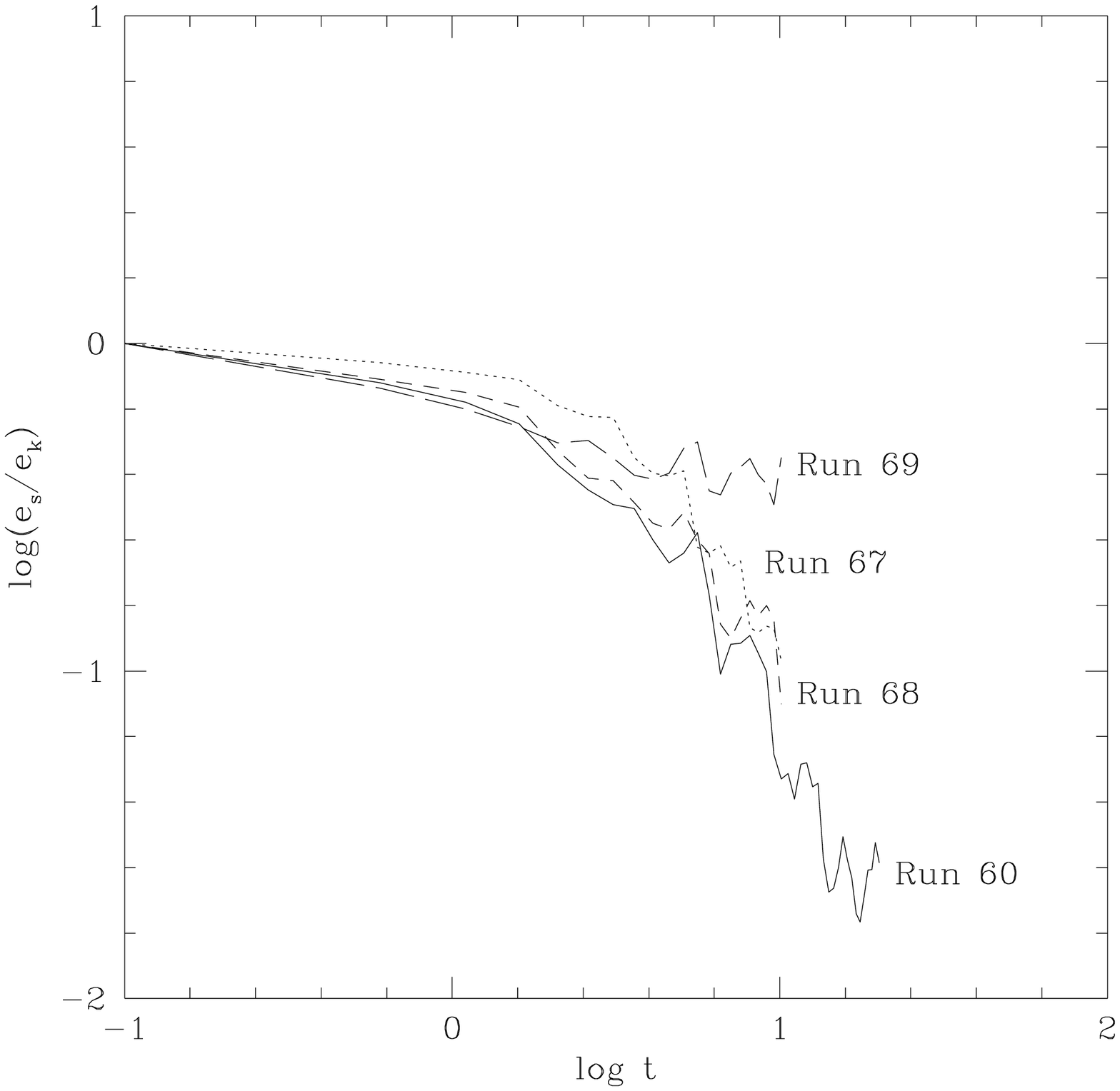,height=2.in,width=2.in}
{\epsfig{file=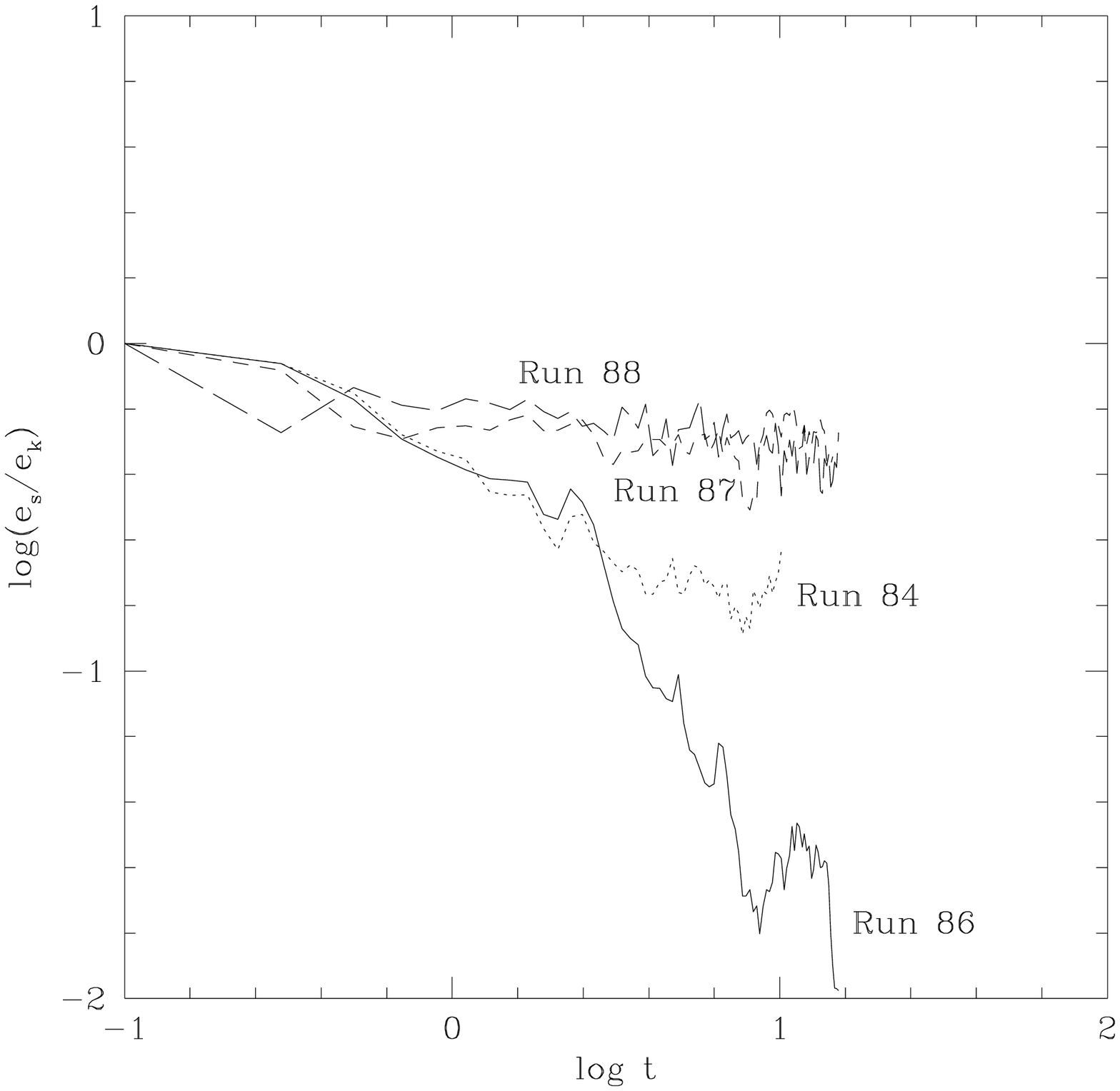,height=2.in,width=2.in}}}
\caption{Evolution of the ratio of rotational to total specific kinetic energy
for various runs. {\it Left}: non-magnetic runs. {\it Right}: magnetic runs.}
\label{eset}
\end{figure}

Fig.\ \ref{eset} ({\it left})
shows the evolution of the ratio $\eset$ of solenoidal to total
kinetic energy per unit mass for four 3D runs. Runs 60 and 67 are \ppm\
runs with ${\game}_{\rm min}= 0.25$ and ${\game}_{\rm min}= 0.75$, both
with fully solenoidal initial velocity modes but fully compressible
forcing ($\Fs \equiv 0$). Run 68
is similar to run 60, but fully thermodynamic. Run 69 is similar to
run 60, but with a Coriolis term added.
In all cases except run 69, $\eset$ is seen to decay at
roughly the same rate in time. Thus :
 
\noindent
i) We have found negligible nonlinear transfer from compressible
to rotational kinetic energies. Similar results in the weakly compressible
case have been found by Kida \& Orszag (1991a,b).
 
\noindent
ii) Additional sources of vorticity, such as the Coriolis force
(acting on large scales), are
necessary for the maintenance of significant amounts of vorticity.
 
Additionally, the presence of an initially uniform magnetic field
also allows for the production of rotational energy from the compressive
motions, with an efficiency proportional to the magnetic field
strength (fig.\ \ref{eset} ({\it right}) ; in runs $86$, $84$, $87$,
$88$,  the non-dimensional field intensity is respectively 
$0.05$, $0.3$, $1$ and $3$). This process is more
important at small scales.

\section*{Flows with Self-Gravity}
 
\subsection*{Non-magnetic case}
 
It can  readily be shown that for a barotropic medium with polytropic
exponent $\game$, the effective Jeans length is :
\begin{equation}
L_{\rm eff} = \Bigl[{\game \pi
c_{\rm i}^2 \over G\rho_\o^{2-\game}} \Bigr]^{1/2} =
\sqrt{{\game}\over{\gamma}} \rho^{{\game-1}\over{2}} L_{\rm J},
\end{equation}
where $c_{\rm i}$ is the isothermal sound speed,  such that $P=c_{\rm
i}^2 \rho$, and $L_{\rm J}$ is the Jeans length based on $c_{\rm i}$.
 
The critical density required for destabilizing a length scale
$L$ is thus $\rho_{\rm J} \propto L^{2/(\game - 2)}$. In order to account
for turbulent compressions acting on $n$ directions,
we consider a volume $V=L^n L_0^{3-n}$, where $L$ is the side of the
volume which varies upon compression, and $L_0$ is the side that
remains unaltered. The critical mass to destabilize this volume is
thus given by $M_{\rm J} \propto L^{n+{2}\over{\game-2}} L_0^{3-n}$.
If $M_{\rm J}$ is a decreasing function of $L$,
turbulent compression can produce gravitationally unstable
structures, i.e, if (see also McKee et al.\ 1993)
\begin{equation}
\game < \gamma_{\rm crit} \equiv 2(1-1/n).
\end{equation}
 
\begin{figure}
\centerline{\epsfig{file=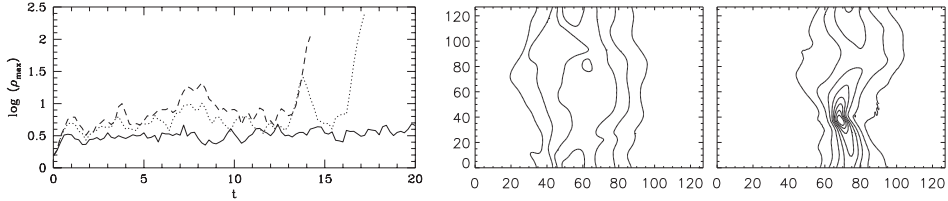,width=5.in}}
\caption{{\it Left}: Evolution of the global density maximum for three runs
with $\game=0.9$ ({\it solid line}), $\game=0.3$ ({\it dotted line}) and
$\game=0.1$ ({\it dashed line}). {\it Right}: Final density fields of two
magnetic simulations, one with $\game=0.9$ ({\it center}) and one with
$\game=0.3$ ({\it right}).}
\label{collapse}
\end{figure}

This result is illustrated in fig.\ \ref{collapse} ({\it left}),
which shows the evolution of the
global density maximum for three single-$\game$ simulations, all with
an isothermal Jeans length $L_{\rm J}=1.1$ times the box size (i.e.,
gravitationally stable) and purely compressible random forcing, but
with $\game=0.9$, 0.3 and 0.1 respectively. While the simulation with
$\game=0.9$ never develops a gravitational collapse, the other two do,
earlier in the case with $\game=0.1$.

\subsection*{Magnetic case}
 
The above result is preserved in the presence of a
magnetic field. Fig.\ \ref{collapse} also
shows contour plots of the final density field in
simulations with an initially uniform magnetic field
along the $x$-direction, a Jeans length $L_{\rm J}=0.9$ (i.e.,
gravitationally unstable with respect to a uniform density
configuration), and purely compressible random forcing. One simulation
has $\game=0.9$ ({\it center}) and the other has $\game=0.3$ ({\it right}). 
The latter is seen to
have undergone collapse, while the former only contracts to a
pancake-type structure supported by thermal pressure against final
collapse. However, for large enough magnetic strengths, we have found
that turbulence
is again not capable of inducing gravitational collapse, suggesting
that a subcritical magnetic regime (Mouschovias \& Spitzer 1976)
cannot be forced to collapse by
large-scale turbulence (although the presence of small-scale turbulent
modes may affect this result).

\end{document}